\documentclass[aps,prb,notitlepage,floatfix,twocolumn,superscriptaddress]{revtex4-2}

\usepackage{amsmath}
\usepackage{amssymb}
\usepackage{mathtools}
\usepackage[tight]{subfigure}
\usepackage{enumitem}
\usepackage{soul}
\usepackage{cancel}
\usepackage{multirow}
\usepackage{tikz}
\usepackage{pifont}
\usepackage{xspace}
\usepackage{comment}
\usepackage{physics}
\usepackage{bbold}
\usepackage{graphicx}
\usepackage{dcolumn}
\usepackage{bm}
\usepackage[breaklinks]{hyperref}
\hypersetup{colorlinks=true, linkcolor=blue, citecolor=blue, filecolor=blue, urlcolor=blue}
\usepackage{ulem}

\begin{document}

\title{Leveraging static quantum many-body scars into period-doubled responses}

\author{Wentai Deng}
\affiliation{School of Physics, Peking University, Beijing 100871, China}

\author{Zhi-Cheng Yang}
\email{zcyang19@pku.edu.cn}
\affiliation{School of Physics, Peking University, Beijing 100871, China}
\affiliation{Center for High Energy Physics, Peking University, Beijing 100871, China}

\date{\today}

\begin{abstract}

We propose a scheme that generates period-doubled responses via periodically driving certain Hamiltonians hosting quantum many-body scars, akin to recent experimental observations in driven Rydberg atom arrays. Our construction takes advantage of an su(2) spectrum generating algebra associated with the static quantum-scarred Hamiltonian, which enacts a $\pi$-rotation within the scar subspace after one period of time evolution with appropriately chosen driving parameters. This yields period-doubled (subharmonic) responses in local observables for any choice of initial state residing in the scar subspace. The quasienergy spectrum features atypical $\pi$-paired eigenstates embedded in an otherwise fully thermal spectrum.
The protocol requires neither a large driving frequency nor a large driving amplitude, and is thus distinct from the prethermalization physics in previous investigations of the driven PXP model. We demonstrate our scheme using several spin-1/2 and spin-1 quantum scarred models possessing an exact su(2) spectrum generating algebra, as well as a symmetry-deformed PXP model, where the su(2) algebra is only approximate. Our results extend the class of models hosting quantum many-body scars that can be leveraged to yield time-crystalline behaviors under periodic driving. 

\end{abstract}

\maketitle

\textit{Introduction.-} Quantum many-body scarring (QMBS) has now been established as a prototypical example of weak ergodicity breaking in quantum many-body systems~\cite{serbyn2021quantum, regnault2022quantum,chandran2023quantum}. In contrast to quantum integrable or many-body localized systems where the full eigenspectrum is nonthermal, quantum many-body scars refer to a small fraction (vanishing in the thermodynamic limit) of nonthermal eigenstates that are embedded in an otherwise fully thermal spectrum. In many cases of interests, these scar states form a tower with almost equal energy spacings that spans the entire many-body spectrum, even at high energies~\cite{turner2018weak, PhysRevB.98.155134}. Therefore, the dynamics starting from certain initial states at high temperatures exhibit nonthermal oscillatory behaviors at late times, as was originally observed on a Rydberg atom quantum simulator~\cite{bernien2017probing}.

Several attempts towards a unified framework for QMBS have been proposed~\cite{PhysRevLett.119.030601, PhysRevB.101.195131, PhysRevB.102.085140, PhysRevLett.126.120604, PhysRevResearch.2.043305, moudgalya2022exhaustive}, many of which encode the scar subspace as an invariant subspace under a higher symmetry than the Hamiltonian itself. In particular, nonabelian symmetries naturally furnish operators satisfying the spectrum-generating algebra~\cite{PhysRevB.101.195131}, giving rise to exact towers of scar states with equidistant energies. 

Recent experiments on Rydberg atom and Bose-Hubbard quantum simulators demonstrate a surprising enhancement of revivals previously observed in static quantum-scarred systems via periodic driving~\cite{bluvstein2021controlling, PhysRevResearch.5.023010}. Moreover, a period-doubled response in local observables was observed for a finite window of driving frequencies, akin to discrete time-crystals~\cite{PhysRevLett.116.250401, PhysRevLett.117.090402, khemani2019brief}. However, unlike the latter, time-crystalline behaviors in driven Rydberg atom arrays only show up using the same initial states that give rise to persistent oscillations in the static case, suggesting an intimate connection to the existence of scars in the undriven system. Theoretical studies of the driven PXP model, which is believed to describe the Rydberg blockade regime, attribute the period-doubled response therein to the emergence of two $\pi$-paired Floquet eigenstates that are superpositions of the N\'eel state and its spatially translated partner~\cite{PhysRevLett.127.090602, PhysRevB.106.104302}.

Unfortunately, due to the complications of the PXP model, an analytical treatment of its driven version is only possible in the case of either a large driving frequencey/amplitude~\cite{PhysRevB.101.245107, PhysRevB.106.104302} or a small deviation from a perfect many-body echo~\cite{PhysRevLett.127.090602}, both giving rise to a prethermalization regime whose relation to the actual experimental protocol is rather obscure. On the other hand, the existence of a vast number of quantum-scarred models with a much cleaner mathematical structure naturally poses the question of whether it is possible to leverage the scar states therein to generate period-doubled dynamics. 

In this work, we propose a scheme that generates period-doubled responses via periodically driving certain quantum-scarred Hamiltonians, thereby extending the quantum scar enabled time-crystalline dynamics to a broader class of models. Our construction takes advantage of an exact su(2) spectrum generating algebra restricted to the scar subspace (a.k.a a quasisymmetry~\cite{PhysRevLett.126.120604}). With appropriately chosen driving parameters, the stroboscopic evolution enacts a $\pi$-rotation within the scar subspace, which gives rise to a subharmonic response in local observables starting from any initial state residing in this subspace. Further inspection of the quasienergy spectrum reveals an $\mathcal{O}(L)$ number of atypical $\pi$-paired eigenstates embedded in an otherwise fully thermal spectrum. The exact nonabelian symmetry within the scar subspace renders the driving protocol analytically tractable, without invoking any perturbative treatment, in contrast to the previously studied PXP model. We demonstrate our scheme using several spin-1/2 and spin-1 quantum-scarred models possessing an exact su(2) spectrum generating algebra, as well as the symmetry-deformed PXP model, where the su(2) algebra is enhanced but still approximate. Interestingly, we find that this example shares certain similarities with the driven PXP model, despite the distinct driving protocols.

\textit{General scheme.-} We start by outlining the general scheme of our construction. Consider the following class of models hosting quantum many-body scars:
\begin{equation}
    H= H_A + J Q^z.
    \label{eq:general}
\end{equation}
The scar subspace $\mathcal{W}$ is annihilated by $H_A$: $H_A |\psi\rangle = 0,\ \forall \ |\psi\rangle \in \mathcal{W}$. We require that this degenerate subspace $\mathcal{W}$ is invariant under an on-site nonabelian symmetry $G$, and yet the full Hamiltonian $H_A$ does not have the $G$ symmetry. In the second term, $Q^z$ is chosen as a (linear superposition of) generator in the Cartan subalgebra of $G$ that lifts the degeneracy of $\mathcal{W}$, with the energy spacing set by $J$. Hereafter, we restrict ourselves to the simplest case where $G=\ $SU(2).
The SU(2) symmetry naturally furnishes raising and lowering operators $Q^{\pm}$ that satisfy a spectrum generating algebra within the subspace $\mathcal{W}$: $[H, Q^{\pm}] \mathcal{W} = \pm J Q^{\pm} \mathcal{W}$. This immediately gives rise to an equally-spaced tower of exact eigenstates of Hamiltonian~(\ref{eq:general}) via repeated actions of $Q^+$ starting from an eigenstate of the Casimir operator and $Q^z$ with eigenvalues $S(S+1)$ and $-S$ in $\mathcal{W}$: $\{|\Omega\rangle, Q^+|\Omega\rangle, \ldots, (Q^+)^{2S}|\Omega\rangle \}$.
Notice that the form of Hamiltonian~(\ref{eq:general}) falls into the category of quasisymmetry construction of QMBS in Ref.~\cite{PhysRevLett.126.120604}. Alternatively, it can be viewed as a special case of the `tunnels to towers' approach discussed in Ref.~\cite{PhysRevResearch.2.043305}. Many known examples of QMBS belong to this class, including the bi-magnon states in the spin-1 $XY$ model~\cite{PhysRevLett.123.147201}, the $\eta$-pairing states in the generalized Hubbard model~\cite{PhysRevB.102.085140}, and the multi-magnon states in the spherical tensor construction~\cite{PhysRevResearch.4.043006} etc.

\begin{figure}[!t]
\includegraphics[width=0.25\textwidth]{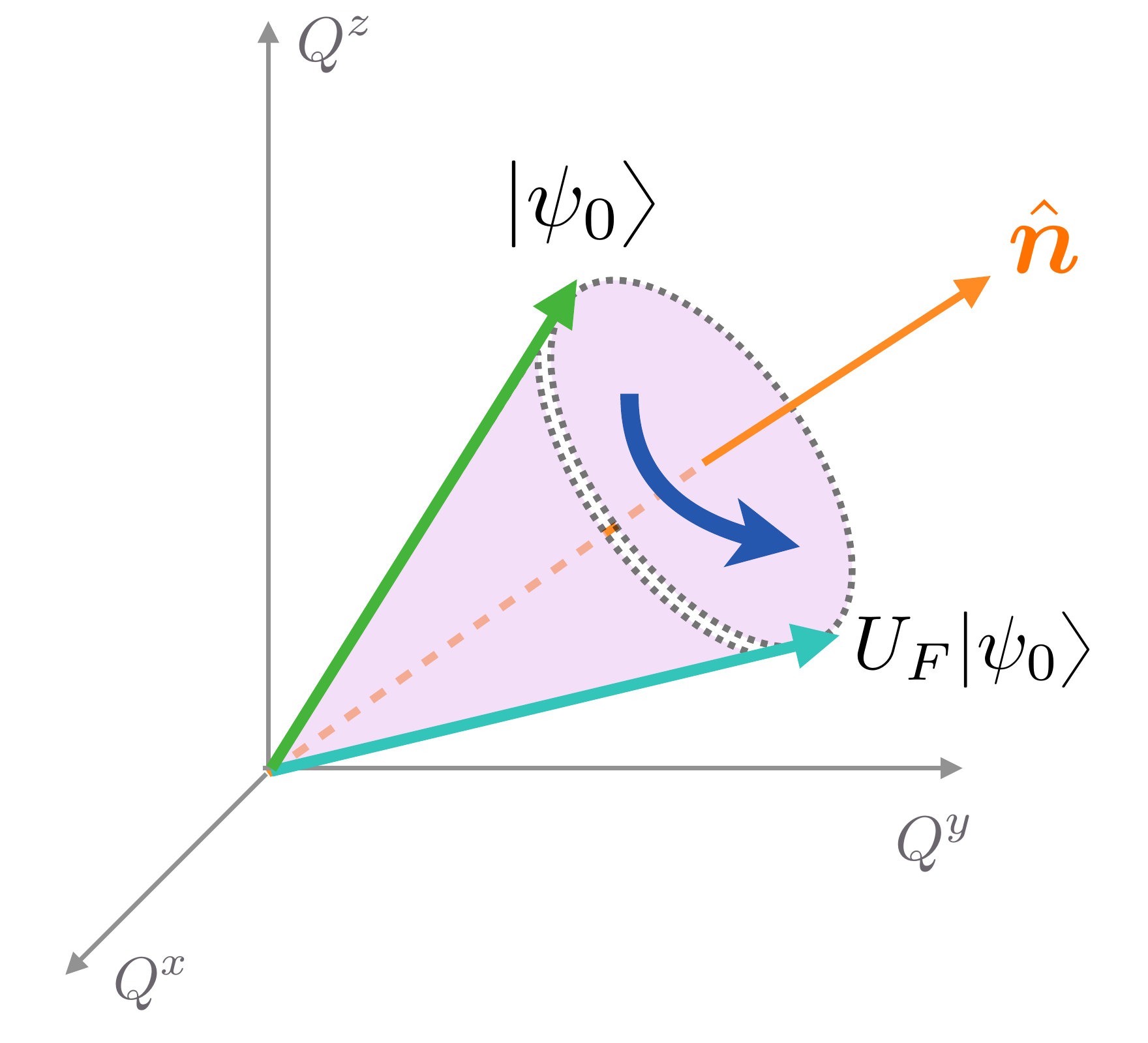}
\caption{Illustration of the stroboscopic dynamics within the scar subspace $\mathcal{W}$. With appropriately chosen driving parameters, the Floquet unitary enacts a $\pi$-rotation about the axis $\hat{{\bm n}}$, giving rise to period-doubled responses in fidelity and local observables.}
\label{fig:rotation} 
\end{figure}

Now, let us consider the following driving protocol:
\begin{equation}
    H(t) = H_A + J Q^z + \lambda(t)\ Q^a,
    \label{eq:floquet_protocol_general}
\end{equation}
where $\lambda(t+T) =\lambda(t)$ with $T$ being the driving period, and $Q^a$ denotes any generator of the su(2) algebra that does not commute with $Q^z$. The Floquet operator after one period of time evolution is given by $U_F = \mathcal{T} \left \{ {\rm exp}[-i \int_0^T H(t) dt] \right \} $, where $\mathcal{T}$ denotes time-ordering. Since $H_A$ annihilates any state within the scar subspace, the Floquet operator projected to the subspace $\mathcal{W}$ takes a particularly simple form:
\begin{eqnarray}
    \mathcal{P}_{\mathcal{W}} U_F \mathcal{P}_{\mathcal{W}} &=& \mathcal{T} {\rm exp} \left \{-i \left[(JT)\ Q^z + \int_0^T dt \ \lambda(t) \ Q^a  \right ]\right \}  \nonumber \\
    &\equiv & {\rm exp} \left[ -i \phi (\hat{{\bm n}}\cdot {\bm Q}) \right],
\label{eq:floquet_general}
\end{eqnarray}
where $\mathcal{P}_{\mathcal{W}}$ denotes projection onto the subspace $\mathcal{W}$.
The Floquet operator effectively generates an SU(2) rotation by an angle $\phi$ within the scar subspace about an axis $\hat{{\bm n}}$, both are determined by the driving parameters. The tower of scar states of the undriven Hamiltonian~(\ref{eq:general}) recombines to form eigenstates along the spin axis $\hat{{\bm n}}\cdot {\bm Q} $, which are eigenstates of $U_F$ with quasienergies $ k\phi $ (mod $2\pi$), $k=-S, -S+1, \ldots, S$. If we choose driving parameters such that the rotation angle $\phi = (2n+1)\pi$ with $n$ integer, the stroboscopic time evolution effectively enacts a $\pi$-rotation within the scar subspace, as illustrated in Fig.~\ref{fig:rotation}. Moreover, the Floquet eigenstates residing in the subspace $\mathcal{W}$ will form two bands with a quasienergy difference of $\pi$ between them. Notice that the simplification of the Floquet unitary into the form~(\ref{eq:floquet_general}) is only true within the subspace $\mathcal{W}$. For a generic choice of $H_A$, the full Floquet operator is quantum chaotic, with the majority of the spectrum featuring an infinite temperature thermal state.

Consider the stroboscopic dynamics starting from an arbitrary initial state $|\psi_0\rangle \in \mathcal{W}$, so long as it is not an eigenstate of $\hat{{\bm n}}\cdot {\bm Q} $. After time evolution of $n$ periods, the state becomes:
\begin{equation}
    |\psi(nT) \rangle = U_F^n |\psi_0\rangle = \sum_{k=-S}^S (-1)^{nk} c_k\ |k\rangle,
\end{equation}
where $|k\rangle$ denotes the Floquet eigenstates and $c_k = \langle k |\psi_0\rangle$. Hence, the state returns to itself only after an even number of periods, which leads to an oscillation in the fidelity of this initial state with twice the period of the drive. Moreover, this subharmonic response also manifests itself in the dynamics of local observables. Consider the expectation value of an operator $Q^b$ that is perpendicular to $\hat{{\bm n}}$: $\hat{{\bm n}}\cdot Q^b = 0$. Then, the effective $\pi$-rotation immediately implies:
\begin{equation}
    \langle Q^b(nT) \rangle = (-1)^n \langle Q^b(0)\rangle,
\end{equation}
which also oscillates with twice the period of the drive, starting from a symmetry-breaking initial state.

\textit{Spin-1/2 chain.-}We demonstrate our general scheme outlined above using concrete examples. Consider the following time-periodic Hamiltonian of a spin-1/2 chain:
\begin{equation}
    H(t) =  J \sum_i \sigma_{i-1}^z \sigma_{i+2}^z P_{i,i+1} \ + \ \Omega \sum_i \sigma_i^x \ +\ \lambda(t) \sum_i \sigma_i^y,
    \label{eq:spin-1/2}
\end{equation}
where $P_{i,i+1}=\frac{1}{4}(1-\vec{\sigma}_i\cdot \vec{\sigma}_{i+1})$ is a projector onto the singlet subspace of two neighboring spins. We choose a three-step square-pulse driving protocol:
\begin{equation}
    \lambda(t) = 
    \begin{cases}
    -\lambda, \quad \quad & 0\leqslant t \leqslant \frac{T}{4} \\
    \lambda, \quad & \frac{T}{4} < t \leqslant \frac{3T}{4}  \\
    -\lambda, \quad   & \frac{3T}{4} < t \leqslant T
    \end{cases}.
\end{equation}
Consider first the undriven Hamiltonian.
The undriven Hamiltonian~(\ref{eq:spin-1/2}) with $\lambda=0$ belongs to a class of models originally considered in Ref.~\cite{PhysRevLett.122.220603}. Since $P_{i,i+1}$ projects onto the singlet subspace, it is then obvious that the $S=\frac{L}{2}$ multiplet of the su(2) algebra is annihilated by the first term of Hamiltonian~(\ref{eq:spin-1/2}), which we identify as the subspace $\mathcal{W}$. This subspace is invariant under an SU(2) quasisymmetry generated by local operators ${\bm S}=\frac{1}{2}\sum_i {\bm \sigma}_i$.
The degeneracy of $\mathcal{W}$ is lifted by the second term of Hamiltonian~(\ref{eq:spin-1/2}), giving rise to an exact tower of $L+1$ scar states labelled by the eigenvalues of the operator $S^x$: $\{|S=\frac{L}{2}, S^x=m_x \rangle,\ m_x = -\frac{L}{2}, \ldots, \frac{L}{2} \}$.

The four-body interactions in the first term of Hamiltonian~(\ref{eq:spin-1/2}) make the system chaotic under periodic driving, which we verify by computing the mean ratio of adjacent quasienergy spacings~\cite{PhysRevB.82.174411}: $\langle r \rangle = \left \langle \frac{{\rm min}(\delta_n, \delta_{n+1})}{{\rm max}(\delta_n, \delta_{n+1})} \right \rangle$, $\delta_n = \phi_n- \phi_{n+1}$, where the quasienergies $\{\phi_n \}$ are rank-ordered in descending order, and we restrict ourselves to the inversion symmetric sector under open boundary condition. We numerically find $\langle r \rangle \approx 0.535$, in agreement with the circular orthogonal ensemble~\cite{PhysRevX.4.041048}. Hence, dynamics starting from generic initial states will quickly thermalize to infinite temperature under the drive. However, the Floquet operator projected to subspace $\mathcal{W}$ is simple:
\begin{eqnarray}
     &{\mathcal P}_{\mathcal W} U_F {\mathcal P}_{\mathcal W } = e^{-i (\Omega \sum_i \sigma_i^x - \lambda \sum_i \sigma_i^y) (T/4)} &  \nonumber \\
    & \times  e^{-i (\Omega \sum_i \sigma_i^x + \lambda \sum_i \sigma_i^y) (T/2)} \ e^{-i (\Omega \sum_i \sigma_i^x - \lambda \sum_i \sigma_i^y) (T/4)}.& \nonumber \\
\end{eqnarray}
The three successive rotations within the scar subspace can be combined into a single one with an angle $\phi$ around some axis $\hat{{\bm n}}$, thereby bringing the Floquet operator to the form of Eq.~(\ref{eq:floquet_general}). $(\phi, \hat{{\bm n}})$ parameterizing the effective net rotation is fully determined by the two phases $\phi_1 \equiv \Omega T$, $\phi_2 \equiv \lambda T$, which can be computed analytically~\cite{SM}:
\begin{equation}
    {\rm cos}\ \frac{\phi}{2} = {\rm cos}^2\ \frac{\varphi}{2} - \frac{\phi_1^2-\phi_2^2}{\phi_1^2+\phi_2^2} \ {\rm sin}^2 \ \frac{\varphi}{2},
    \label{eq:spin-1/2_analytic}
\end{equation}
where $\varphi \equiv \sqrt{\phi_1^2+\phi_2^2}$. Thus, one can choose driving parameters such that $\phi=\pi$ in the above equation, which gives rise to period-doubled responses starting from an initial state within $\mathcal{W}$.

\begin{figure}[!t]
\includegraphics[width=0.5\textwidth]{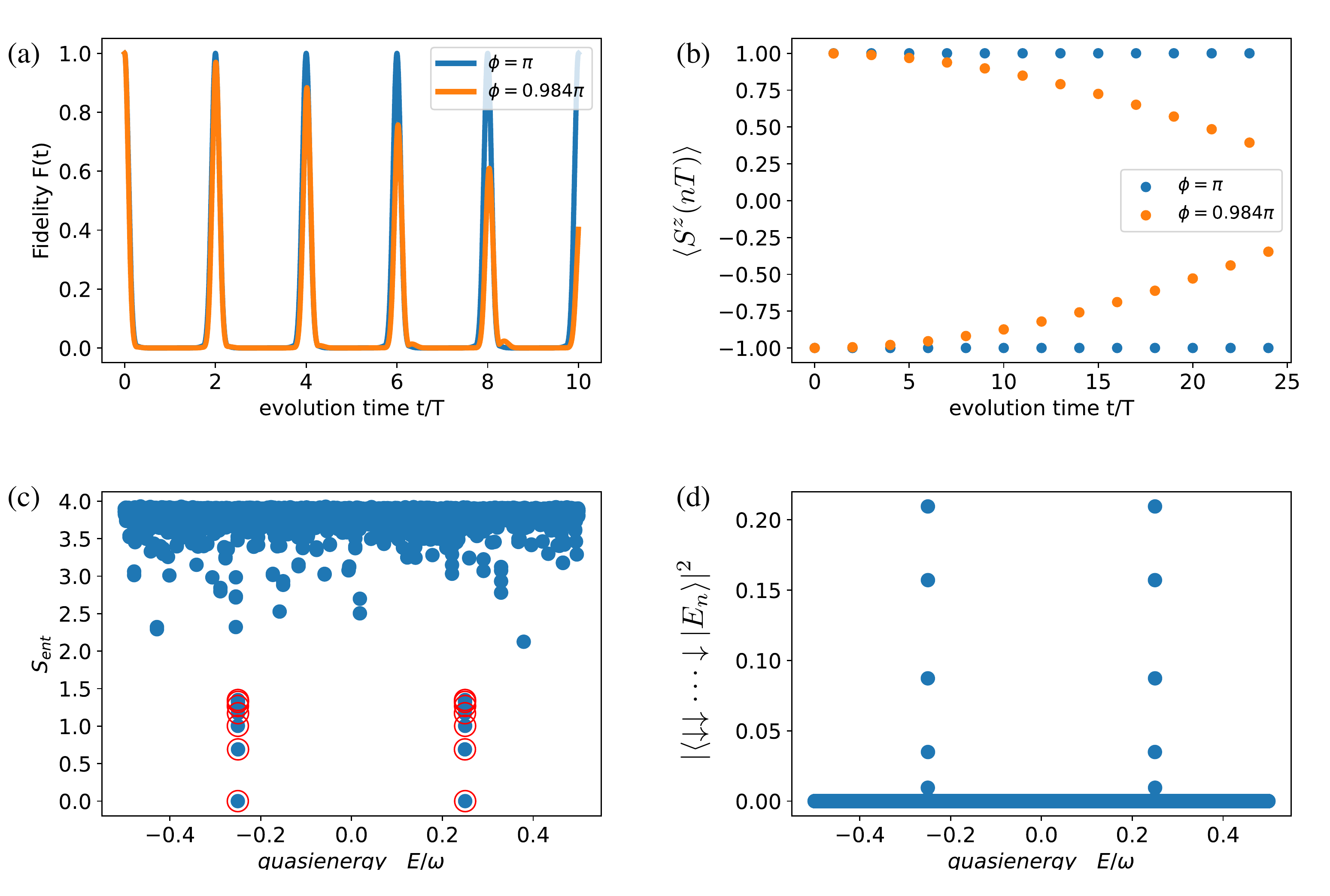}
\caption{Numerical results for the driven spin-1/2 model in Eq.~(\ref{eq:spin-1/2}). (a) The fidelity $F(t) = |\langle \psi_0|\psi(t)\rangle|^2$ as a function of time, which exhibits a perfect period-doubled oscillation for an effective rotation angle $\phi=\pi$ within the scar subspace. The oscillation becomes damped when $\phi$ deviates from $\pi$. (b) Expectation value of the total $z$-magnetization at stroboscopic times $\langle S^z(nT)\rangle$, which exhibits perfect and damped period-doubled oscillations similarly to the fidelity. (c) Entanglement entropies of the Floquet eigenstates at $\phi=\pi$, where states within subspace $\mathcal{W}$ are circled in red. The spectrum features a set of $\pi$-paired eigenstates within $\mathcal{W}$ with low entanglement. (d) Overlap between the initial state $|\psi_0\rangle = |\downarrow \downarrow \cdots \downarrow\rangle$ and the Floquet eigenstates. We use $L=13$, $\Omega T \approx 1.8$ and $\lambda T \approx 1.5066$ for $\phi=\pi$, and $\Omega T \approx 1.8$ and $\lambda T \approx 1.5966$ for $\phi=0.984\pi$.} 
\label{fig:spin_1/2} 
\end{figure}

We verify our analytical predictions numerically, as shown in Fig.~\ref{fig:spin_1/2}. We choose as our initial state a fully polarized state along the $z$ direction: $|\psi_0\rangle = |\downarrow \downarrow \cdots \downarrow\rangle$, which corresponds to the lowest weight state in the $S=\frac{L}{2}$ multiplet. For a choice of driving parameters $(\Omega T, \lambda T)$ such that $\phi = \pi$, we find that the fidelity of the initial state as a function of time $F(t)= |\langle \psi_0|\psi(t)\rangle|^2$ indeed exhibits perfect revivals with twice the period of the drive [Fig.~\ref{fig:spin_1/2}(a)]. This subharmonic response also manifests itself in the expectation value of observables, namely, the total $z$-magnetization at stroboscopic times, as shown in Fig.~\ref{fig:spin_1/2}(b). When $\phi$ slightly deviates from $\pi$, rivivals of the fidelity and magnetization are no longer perfect [Fig.~\ref{fig:spin_1/2}(a)\&(b)]. However, since in this case the dynamics still corresponds to rotations of a large `spin' in the subspace $\mathcal{W}$, this damping is not due to a leakage of the quantum state outside of $\mathcal{W}$. As we show in Supplemental Material (SM)~\cite{SM}, at longer times the fidelity exhibits rather complicated quasiperiodic oscillations with a slowly varying amplitude modulation on top of the fast oscillations. When the deviation from $\pi$ is small, the fidelity still oscillates at approximately twice the period of the drive with a gradually decaying amplitude within a moderate time window. 
This suggests that one does not have to fine tune the value of $\phi$ in order to see signatures of period-doubled responses in finite systems. To relate the observed subharmonic responses to properties of the Floquet eigenstates, in Fig.~\ref{fig:spin_1/2}(c)\&(d) we plot the bipartite entanglement entropies of the Floquet eigenstates under a bipartitioning of the system in the middle, and the overlap of $|\psi_0\rangle$ with the eigenstates, respectively. We find $(L+1)/2$ pairs (with $L$ odd) of $\pi$-paired eigenstates residing completely within $\mathcal{W}$ with low entanglement. The majority of the Floquet eigenstates are close to maximally entangled, confirming the quantum chaotic nature of the full Floquet unitary $U_F$. Thus, the model features a special subset of $\pi$-paired eigenstates enabled by the existence of scars in the undriven Hamiltonian, which leads to time-crystalline behaviors for certain initial states.


The protocol can be straightforwardly generalized to yield oscillations in fidelity and local observables with a period equal to other integer multiple of the driving period $nT$, by setting $\phi = \frac{2\pi}{n}$ in Eq.~(\ref{eq:spin-1/2_analytic})~\cite{SM}. The same approach also works for other quantum-scarred models that fall into the category of Eq.~(\ref{eq:floquet_general}) with an su(2) spectrum generating algebra. In the SM~\cite{SM}, we give another example of the spin-1 $XY$ model hosting an exact tower of scars with an su(2) algebra~\cite{PhysRevLett.123.147201}.

\begin{figure}[!t]
\includegraphics[width=0.48\textwidth]{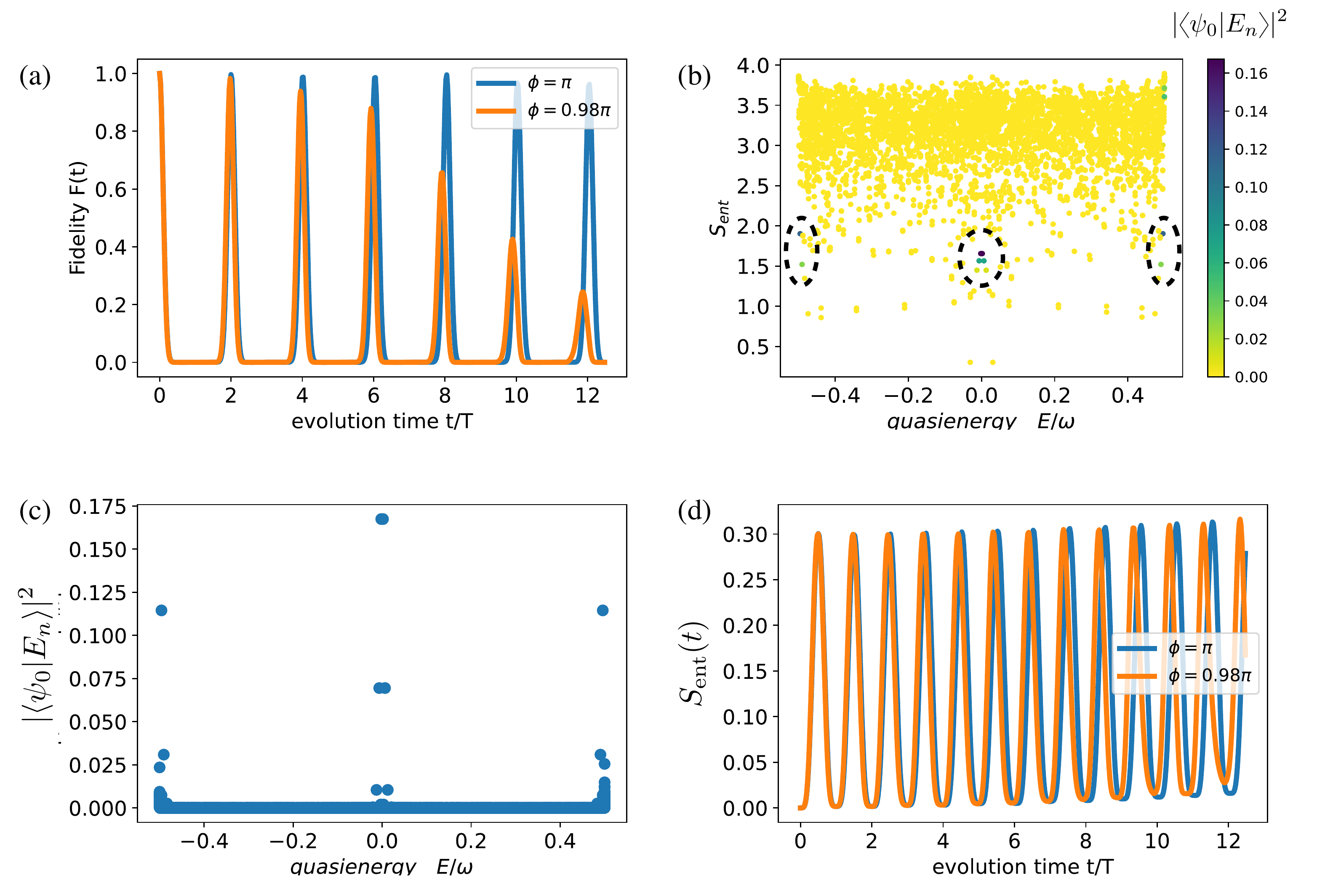}
\caption{Numerical results for the driven deformed-PXP model in Eq.~(\ref{eq:pxp_deform}). (a) The fidelity $F(t) = |\langle \psi_0|\psi(t)\rangle|^2$ as a function of time, which exhibits period-doubled oscillations for an effective rotation angle $\phi=\pi$ and $\phi=0.98\pi$ within the scar subspace. (b) Bipartite entanglement entropies of the Floquet eigenstates at $\phi=\pi$, where the overlaps with the initial state $|\psi_0\rangle = |\downarrow \uparrow \downarrow \uparrow \cdots \downarrow \uparrow \rangle$ are color-coded. (c) Overlap between the N\'eel initial state and the Floquet eigenstates. (d) Time evolution of the entanglement entropy starting from $|\psi_0\rangle$. We use $L=18$, $\lambda T \approx 2.1347 $ for $\phi=\pi$, and $\lambda T \approx 2.1867$ for $\phi=0.98\pi$.}
\label{fig:pxp} 
\end{figure}

\textit{Deformed PXP model.-} We move on to a model where the su(2) algebra associated with the scar subspace is only approximate. The model we shall consider is based on a particular deformation of the PXP model that leads to an improved su(2) algebra within the scar subspace and enhanced quantum revivals~\cite{PhysRevB.101.165139}. Consider the following Hamiltonian:
\begin{eqnarray}
    H(t) &=&  \sum_i P_{i-1} \sigma_i^x P_{i+1}  \nonumber \\
    && + \ \delta \sum_i (P_{i-2} P_{i-1} \sigma_i^x P_{i+1} + P_{i-1} \sigma_i^x P_{i+1} P_{i+2}) \nonumber \\
    && + \ \lambda(t) \sum_i (n_{2i} - n_{2i-1}) \nonumber \\
    &\equiv& H_{PXP} + H_{\rm deform} + H_{\rm drive}.
    \label{eq:pxp_deform}
\end{eqnarray}
The first term is the usual PXP model, where $P_i = |0\rangle \langle 0|_i$ is a projector on the empty state on site $i$, and we use the spin and particle descriptions for the basis states interchangeably $|0\rangle =|\downarrow\rangle$, $|1\rangle =|\uparrow \rangle$. This term guarantees that no adjacent sites can be simultaneously excited to the Rydberg state $|1\rangle$ due to Rydberg blockade. The second term is the deformation originally found in Ref.~\cite{PhysRevB.101.165139} that leads to an embedding of a scar subspace $\mathcal{W}$ as a representation of an approximate su(2) algebra at an optimal value of $\delta \approx 0.108$. The last term is a periodic drive of a staggered chemical potential on two sublattices. For simplicity, we consider a two-step square-pulse driving protocol: $\lambda(t) = -\lambda$ for $0\leq t < \frac{T}{2}$, and $\lambda(t) = \lambda$ for $\frac{T}{2} \leq t < T$. Notice that the drive considered above is different from that used experimentally for the unperturbed PXP model~\cite{bluvstein2021controlling, PhysRevB.106.104302}. In that case, the system is driven by a time-dependent \textit{uniform} chemical potential, instead of a staggered one. Our choice of a staggered chemical potential is motivated by an associated su(2) algebra, which we describe below.

The connection of this model to the general form of Eq.~(\ref{eq:floquet_protocol_general}) can be made explicit by using an effective spin-1 description of the PXP model~\cite{PhysRevA.107.023318}. Define the block-spin basis states by grouping two adjacent sites: $|0\rangle_i \equiv |\downarrow \downarrow \rangle_{2i-1, 2i}$, $|-\rangle_i \equiv |\uparrow \downarrow \rangle_{2i-1,2i}$, $|+\rangle_i \equiv |\downarrow \uparrow \rangle_{2i-1,2i} $. The block-spin then can be viewed as an effective spin-1 degree of freedom, in terms of which the PXP model takes a simple form: $H_{PXP} = \sqrt{2} \sum_{i} S_i^x + H'$, where $S_i^x$ is the spin-1 operator on bond $i \equiv (2i-1, 2i)$, and $H'$
forbids the configuration $|+-\rangle$ from being generated under the dynamics. The particular form of $H'$ is not needed for our purpose, but can be worked out in a straightforward way. It was numerically demonstrated in Ref.~\cite{PhysRevA.107.023318} that the spectrum of $H_{PXP}+H_{\rm deform}$ contains a tower of scar states with an equal energy spacing of approximately $\sqrt{2}$. One can thus identify $J Q_z \equiv \sqrt{2} \sum_i S_i^x$, $H_A \approx H' + H_{\rm deform}$, and $\lambda(t) Q^a \equiv \lambda(t) \sum_i (n_{2i-1} - n_{2i}) = \lambda(t) \sum_i S_i^z $, by comparing with Eq.~(\ref{eq:floquet_protocol_general}). Therefore, the model we consider indeed falls into the category of the general form~(\ref{eq:floquet_protocol_general}), but with an approximate su(2) algebra associated with the subspace $\mathcal{W}$.

The spin-1 representation of Hamiltonian~(\ref{eq:pxp_deform}) also makes it possible to derive an analytic expression for the effective SU(2) rotation angle $\phi$ within the scar subspace in terms of the driving parameters, neglecting the effect of su(2)-breaking. The precise expression is quite lengthy, and we show it in the SM~\cite{SM}. In Fig.~\ref{fig:pxp}(a), we find that, although the su(2) algebra is only approximate within the subspace $\mathcal{W}$, choosing a driving parameter that gives $\phi=\pi$ still produces nearly perfect period-doubled revivals in fidelity, starting from the N\'eel initial state $|\psi_0\rangle = |\downarrow \uparrow \downarrow \uparrow \cdots \downarrow \uparrow \rangle$. This provides another evidence that Hamiltonian~(\ref{eq:pxp_deform}) indeed secretly falls into the general class of Eq.~(\ref{eq:floquet_general}).
The oscillation can similarly be attributed to the emergence of approximately $\pi$-paired Floquet eigenstates residing predominantly in $\mathcal{W}$ with low entanglement, as shown in Fig.~\ref{fig:pxp}(b)\&(c). Fig.~\ref{fig:pxp}(d) further plots the entanglement entropy as a function of time, which oscillates within a window of small value, again indicating that the dynamics is mostly constrained within the scar subspace. However, since the su(2) algebra is only approximate in this case, the entanglement entropy exhibits persistent oscillations around an average value that gradually increases with time (although very slowly), which reflects the leakage of the initial state outside the subspace $\mathcal{W}$~\cite{SM}. At very short times, the entanglement entropy returns to nearly zero after one driving period. Interestingly, we find that the state evolves into the spatially translated partner of the N\'eel initial state $|\uparrow \downarrow \uparrow \downarrow \cdots \uparrow \downarrow \rangle$ at that point  (with an overlap of 0.999 for $L=18$). That the state oscillates between the two N\'eel states at stroboscopic times is reminiscent of the driven PXP model, although the driving protocols are rather different in these two cases.

\textit{Discussion.-} In this work, we propose a simple protocol that leverages an algebraic structure present in many quantum-scarred Hamiltonians to generate time-crystalline behaviors. In particular, we consider situations where the scar subspace is invariant under an SU(2) symmetry, whose generators can be coupled to a time-periodic drive that enacts a $\pi$-rotation (or other integer fractions of $2\pi$) within the scar subspace. Dynamics starting from any initial state within the static scar subspace will exhibit period-doubled oscillations, due to a spectral pairing of scar states by a quasienergy $\pi$. We demonstrate our protocol using several models with an exact su(2) spectrum generating algebra, as well as the deformed PXP model where the su(2) algebra is only approximate. We remark that our scheme directly takes advantage of static scars in the undriven system, and hence distinct from intrinsic Floquet scars in the literature~\cite{PhysRevLett.129.133001}.

An interesting open question is whether our protocol for the deformed PXP model can be related to the experimentally observed period-doubled phenomena in the unperturbed PXP model, although in the latter case the drive couples to a uniform chemical potential rather than a staggered one. Generalizations of this protocol to cases where the scar subspace has a higher rank Lie group symmetry, or the scar subspace does not form an irrep of the su(2) algebra (e.g. the exact tower of scars in the AKLT model~\cite{PhysRevResearch.2.043305}) are also outstanding questions that we leave for future work.

\textit{Note added.-} While finishing up this work, we became aware of an independent work exploring a similar idea in a different setup, which will appear in the same arXiv posting~\cite{note}.

\textit{Acknowledgments.-} Z.-C.Y. is supported by a startup fund at Peking University. Numerical simulations were performed on High-performance Computing Platform of Peking University.

\bibliography{reference}

\newpage
\onecolumngrid
\appendix

\subsection*{Supplemental Material for ``Leveraging static quantum many-body scars into period-doubled responses"}

\section{Derivation of the driving parameters for the spin-1/2 model}

We derive Eq. (9) in the main text relating the rotation angle $\phi$ within the scar subspace to the driving parameters. To simplify our calculation, we define a new Floquet operator which is related to the one defined in Eq.(8) by a time translation $t\rightarrow t+\frac{T}{4}$:
\begin{equation}
     {\mathcal P}_{\mathcal W} \widetilde{U}_F {\mathcal P}_{\mathcal W } = e^{-i (\Omega \sum_i \sigma_i^x + \lambda \sum_i \sigma_i^y) (T/2)} \ e^{-i (\Omega \sum_i \sigma_i^x - \lambda \sum_i \sigma_i^y) (T/2)}.
\end{equation}
This amounts to a gauge choice of the initial point $t_0$ in defining the Floquet unitary, which does not affect the quasienergy spectrum~\cite{d2016quantum}. Define dimensionless variables $\phi_1\equiv \Omega T$ and $\phi_2 \equiv \lambda T$, the projected Floquet unitary can be written as:
\begin{equation}
    {\mathcal P}_{\mathcal W} \widetilde{U}_F {\mathcal P}_{\mathcal W } = \prod_k e^{-i \hat{{\bm n}}_2 \cdot {\bm \sigma}_k \varphi/2} e^{-i \hat{{\bm n}}_1 \cdot {\bm \sigma}_k \varphi/2} \equiv \prod_k e^{-i \hat{{\bm n}} \cdot {\bm \sigma}_k \phi /2}, 
\end{equation}
where the rotation axes $\hat{{\bm n}}_1 = (n_x, -n_y, 0)$, $\hat{{\bm n}}_2 = (n_x, n_y, 0)$, $n_x = \phi_1/\sqrt{\phi_1^2+\phi_2^2}$, $n_y = \phi_2/\sqrt{\phi_1^2+\phi_2^2}$, and the rotation angle within each step $\varphi= \sqrt{\phi_1^2+\phi_2^2}$. Expanding the above equation on both sides using the identity $e^{-i \hat{{\bm n}} \cdot {\bm \sigma}\theta} = {\rm cos}\theta - i (\hat{{\bm n}}\cdot {\bm \sigma}) {\rm sin}\theta$, and matching the part proportional to the identity, we obtain:
\begin{equation}
    {\rm cos}\ \frac{\phi}{2} = {\rm cos}^2\ \frac{\varphi}{2} - (n_x^2-n_y^2) \ {\rm sin}^2 \ \frac{\varphi}{2},
\end{equation}
which is Eq. (9) in the main text.

\section{Additional numerical results on the spin-1/2 model}

In this section, we provide additional numerical results on the spin-1/2 model studied in the main text, which include 
(1) oscillations of fidelity and local observables in cases of imperfect rotations at longer times; (2) period-tripled dynamics upon choosing $\phi=\frac{2\pi}{3}$.

\begin{figure}[hbt]
  \centering
  \subfigure[]{
    \includegraphics[width=0.45\textwidth]{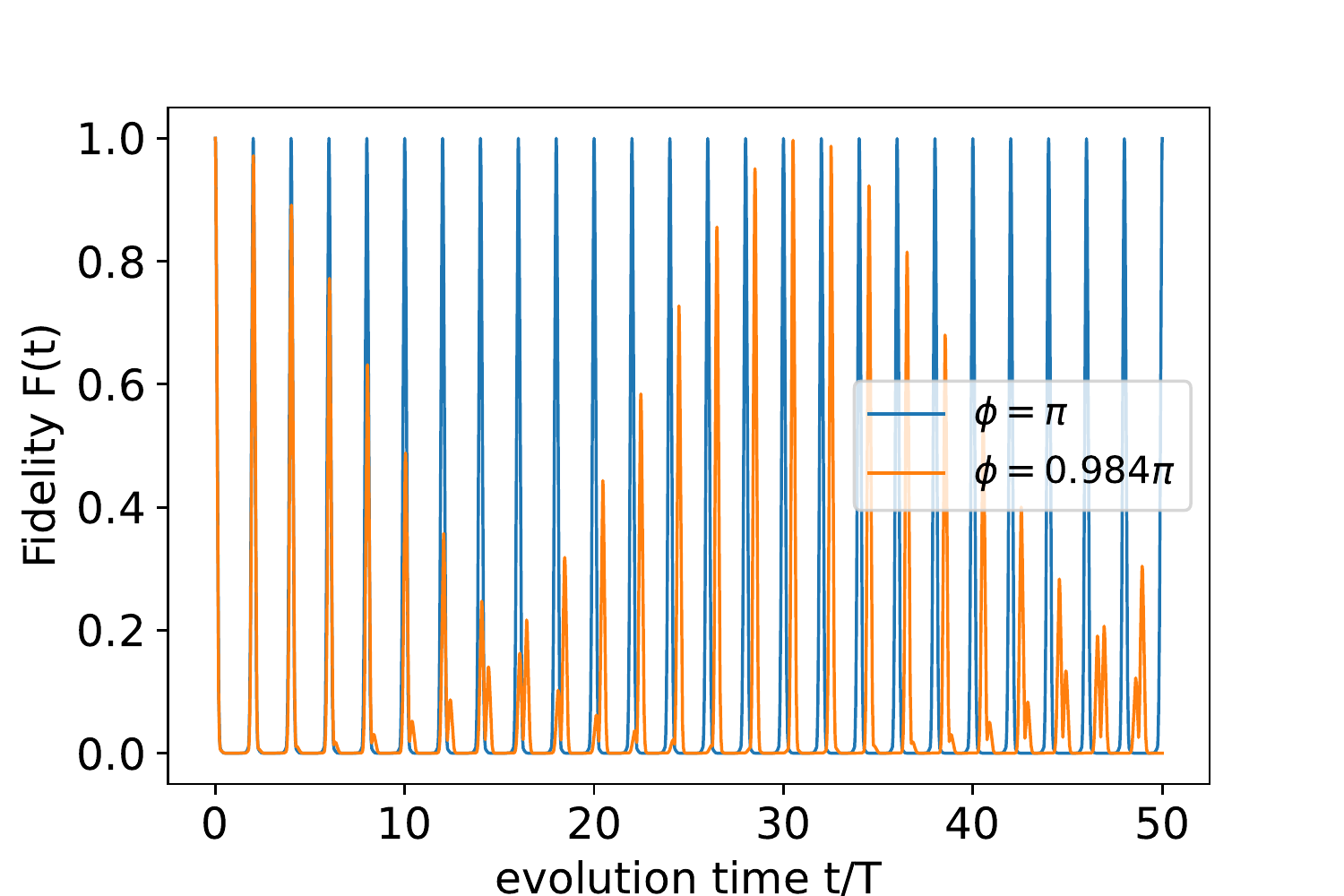}
    \label{fig:1S}
  }%
  \subfigure[]{
    \includegraphics[width=0.45\textwidth]{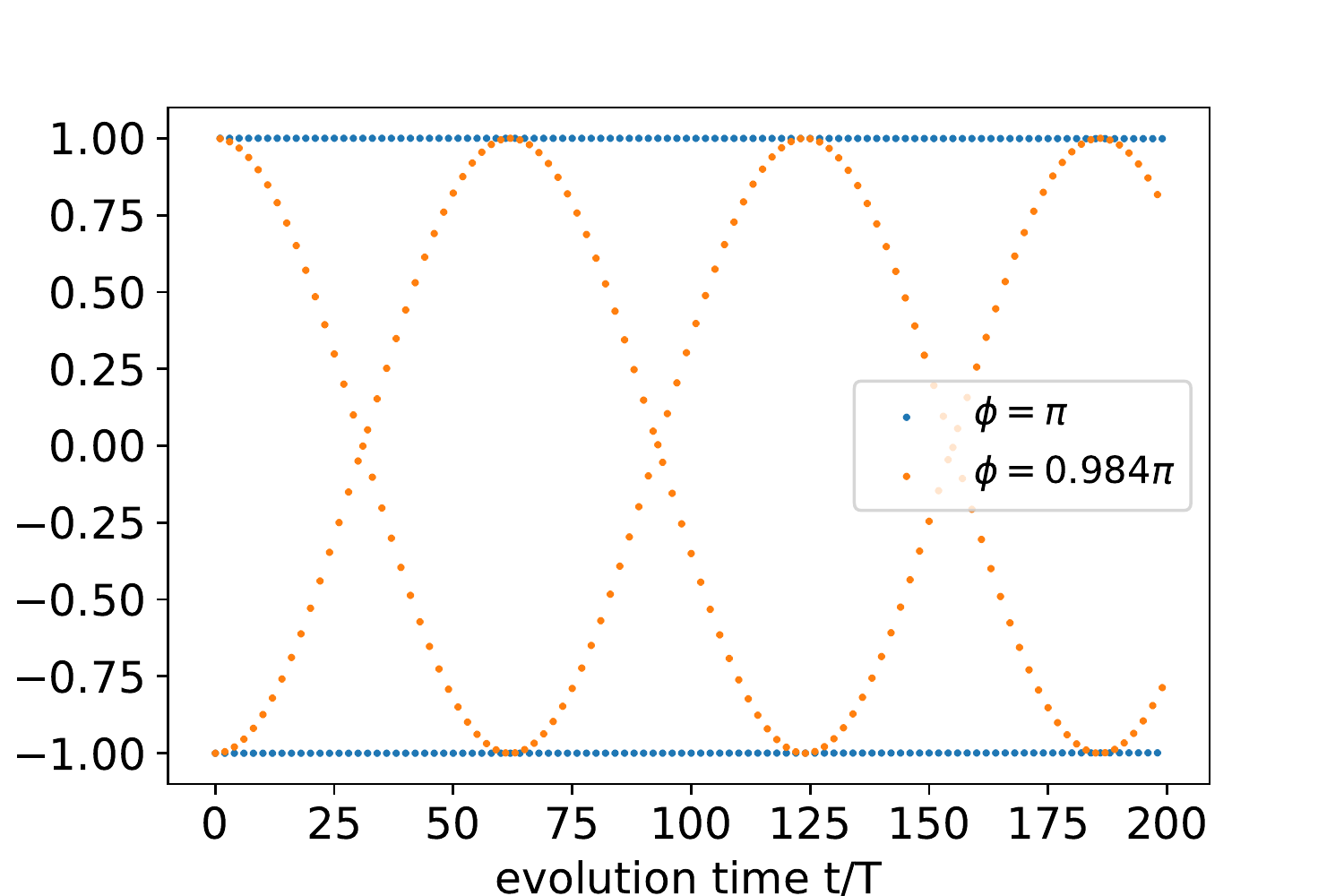}
    \label{fig:1S}
  }%
  \caption{Evolution of (a) the fidelity $F(t)$ and (b) magnetization $\langle S^z(nT)\rangle$ for the driven spin-1/2 model at longer times. The choice of parameters is the same as in Fig. 2 of the main text. }
  \label{fig:spin_1/2_additional}
\end{figure}

\begin{figure}[hbt]
  \centering
  \subfigure[]{
    \includegraphics[width=0.33\textwidth]{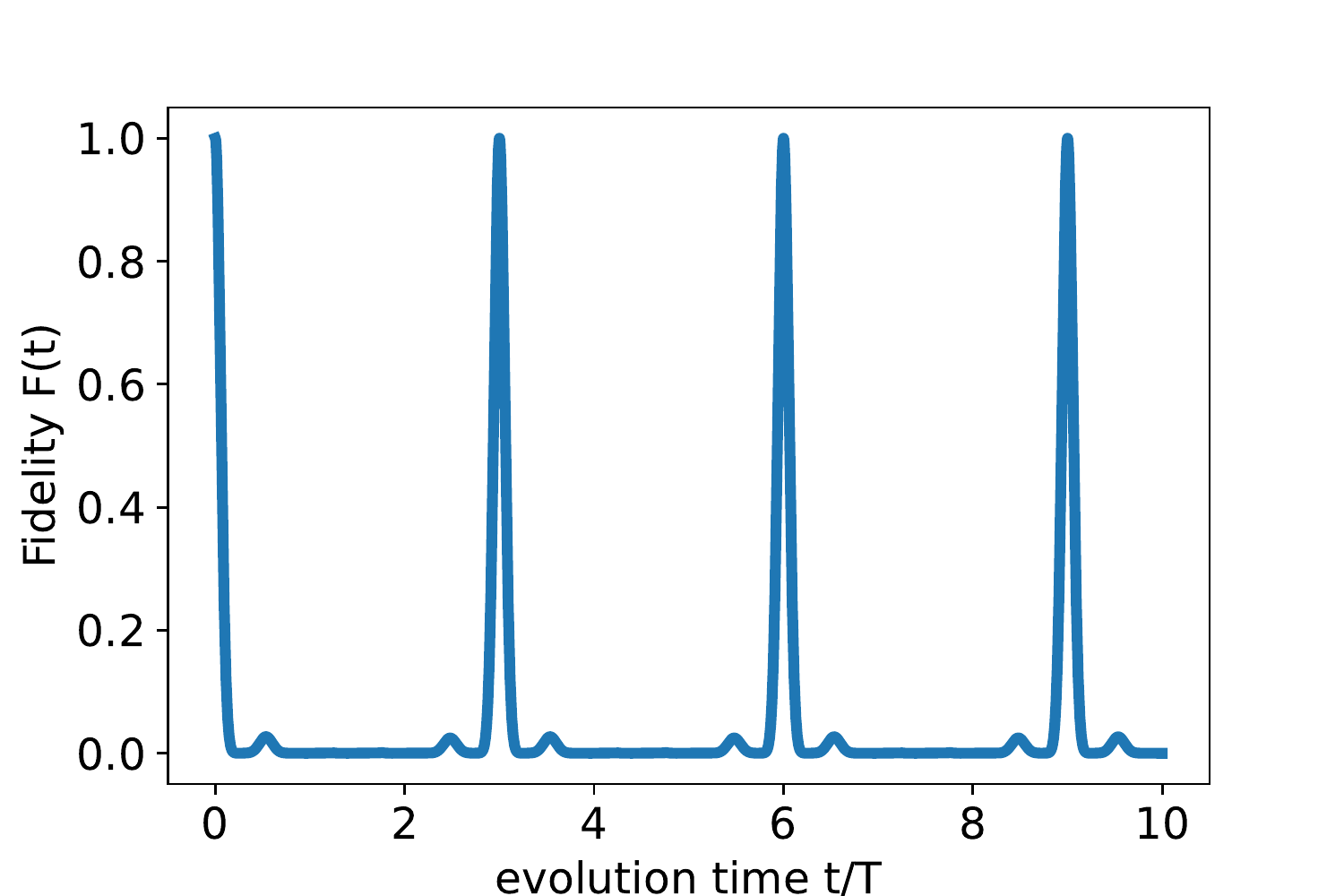}
    \label{fig:1S}
  }%
  \subfigure[]{
    \includegraphics[width=0.33\textwidth]{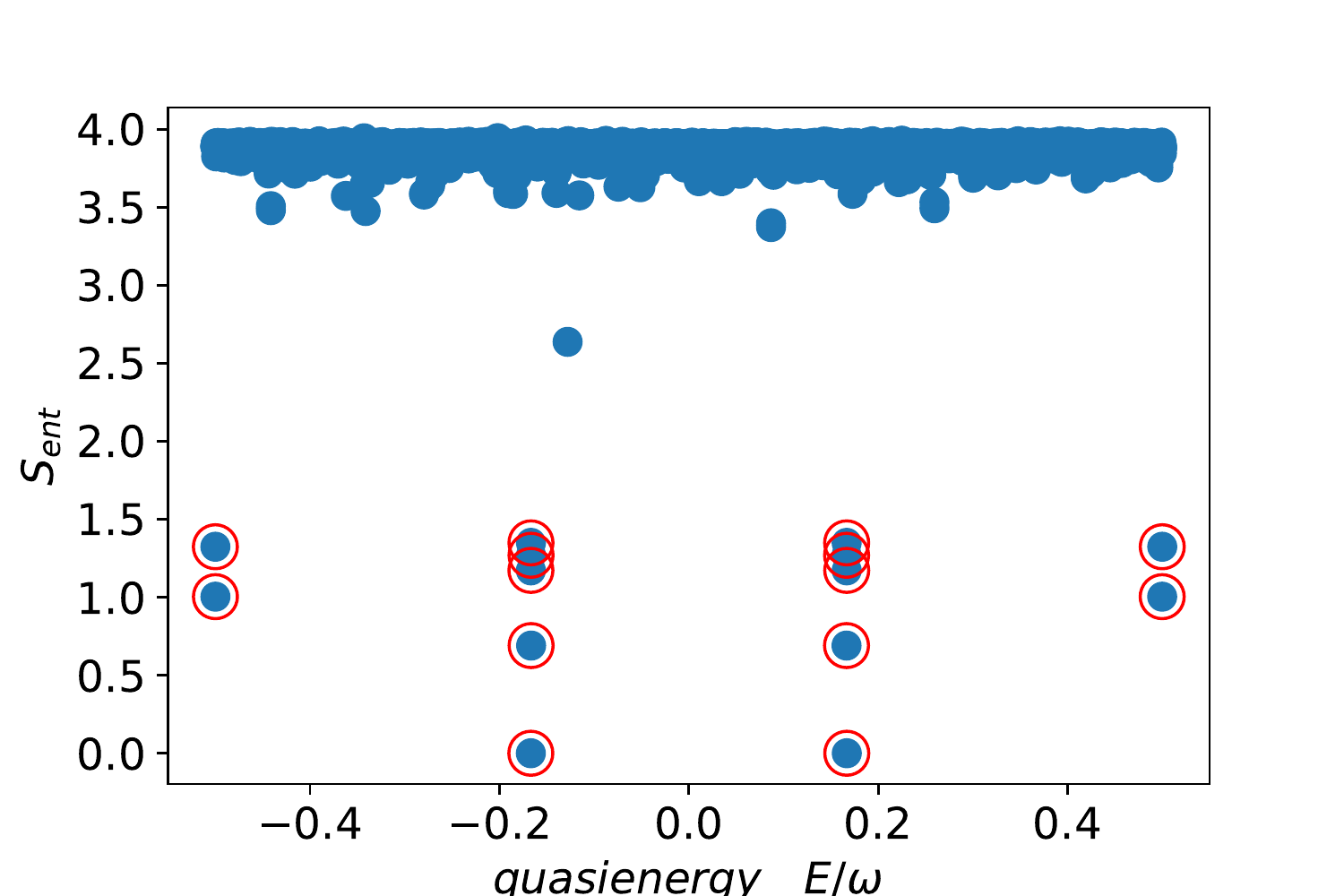}
    \label{fig:1S}
  }%
  \subfigure[]{
    \includegraphics[width=0.33\textwidth]{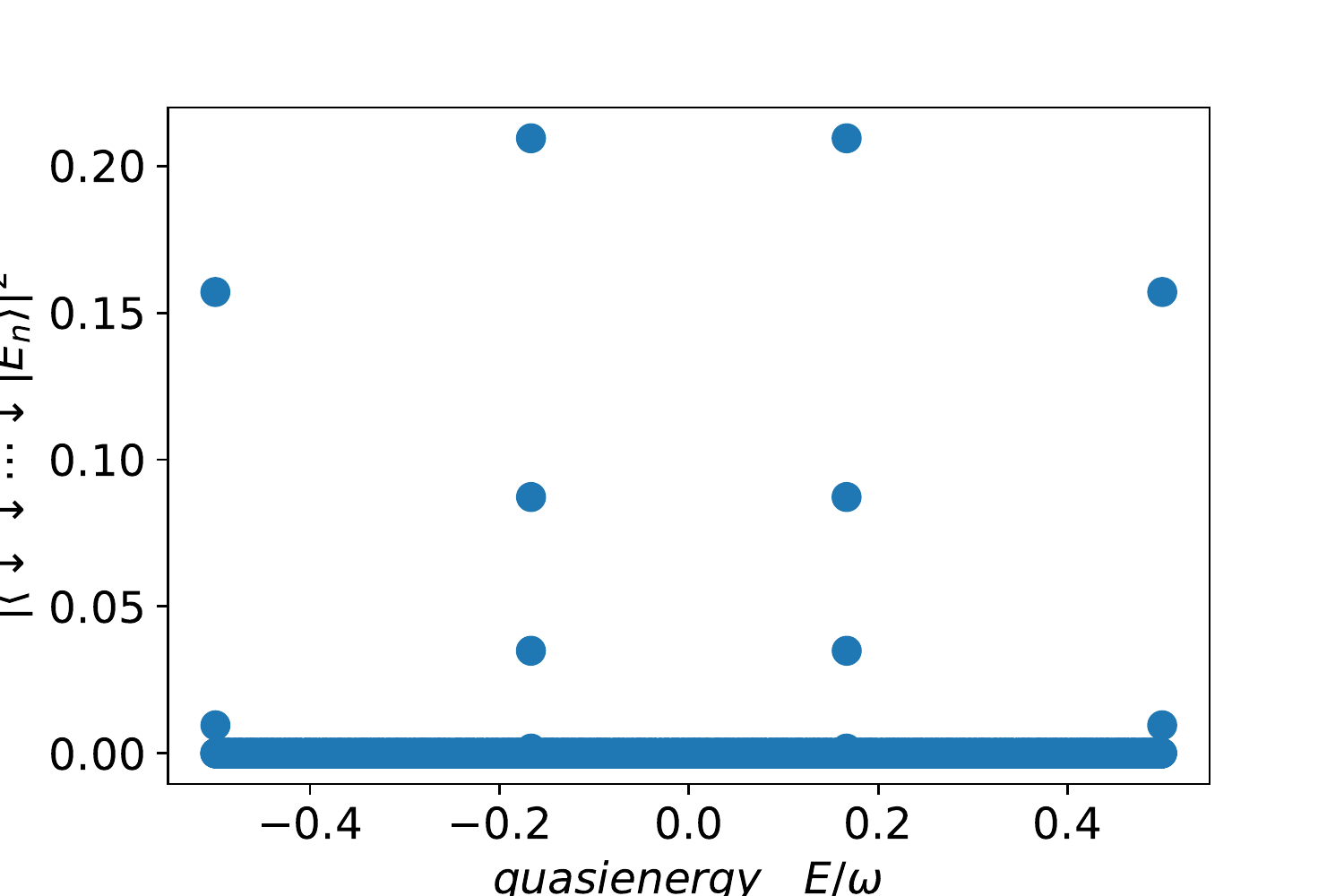}
    \label{fig:2S}
  }
  \caption{Numerical results for the spin-1/2 model with a different set of driving parameters corresponding to $\phi=\frac{2\pi}{3}$. (a) The fidelity $F(t)$ exhibits perfect revivals at integer multiples of $3T$. (b) The quasienergy spectrum features a set of weakly-entangled eigenstates with quasienergy spacing $\frac{2\pi}{3}$ between them. (c) Overlap of the initial state $|\psi_0\rangle= |\downarrow \downarrow \cdots \downarrow\rangle$ with the Floquet eigenstates. Results are obtained for system size $L=13$, $\Omega T \approx 1.8$, and $\lambda T \approx 3.0384 $, such that $\phi=\frac{2\pi}{3}$.}
  \label{fig:triple}
\end{figure}

In Fig.~\ref{fig:spin_1/2_additional}, we show the evolution of the fidelity and magnetization for an imperfect rotation angle at longer times. We find that the fidelity now exhibits rather complicated quasiperiodic oscillations with a slowly varying amplitude modulation on top of the approximately subharmonic oscillation. This is the generic behavior for an imperfect rotation angle that is an irrational fraction of $2\pi$. Consider a small deviation of $\phi$ from $\pi$: $\phi = (1-\epsilon)\pi$. The stroboscopic dynamics now becomes:
\begin{equation}
    |\psi(nT)\rangle = \sum_{k=-S}^{S} e^{-i(1-\epsilon)\pi nk} c_k |k\rangle = \sum_{k=-S}^{S} (-1)^{nk} e^{i\epsilon n k \pi} c_k |k\rangle.
\end{equation}
At short times $\epsilon n L \ll 1$, the fidelity oscillates at approximately twice the period of the drive, with a gradually decaying amplitude. At longer times when $\epsilon n L \approx 1$, the fidelity exhibits a more complicated structure. In particular, the peaks of $F(t)$ no longer coincide with stroboscopic times. In fact, one finds that $F(t)$ is close to minimum at $t=nT \gtrsim 20$ in Fig.~\ref{fig:spin_1/2_additional}(a). This explains why the dynamics of $S^z$ in Fig.~\ref{fig:spin_1/2_additional}(b) appears to be out of synchronization with that of the fidelity.

Secondly, we show in Fig.~\ref{fig:triple} that our protocol is also capable of generating period-tripled oscillations upon tuning the driving parameters such that $\phi=\frac{2\pi}{3}$. We find that the fidelity of the initial state $|\psi_\rangle = |\downarrow \downarrow \cdots \downarrow \rangle$ now oscillates with a period equal to three times the driving period. Such a period-tripled dynamics can again be attributed to the spectral pairing property of the Floquet quasienergies, as shown in Fig.~\ref{fig:triple}(b). We find that quasienergy spectrum contains a set of weakly entangled eigenstates embedded in an otherwise fully thermal spectrum (with bipartite entanglement entropies being close to the Page value). These special eigenstates are equally spaced in quasienergy by $\frac{2\pi}{3}$, as expected. The initial state again resides completely within the subspace of these states [Fig.~\ref{fig:triple}(c)].

\section{Analytic form of the rotation angle $\phi$ for the deformed PXP model}
\label{sec:PXP}

We derive an analogous expression of Eq. (9) in the main text for the deformed PXP model that relates the rotation angle $\phi$ within the scar subspace to the driving parameters. We consider the time-periodic Hamiltonian:
\begin{equation}
    H(t) = \Omega \ H_{PXP} + H_{\rm deform} + \lambda(t) \sum_i (n_{2i-1} - n_{2i}),
\end{equation}
with
\begin{equation}
    \lambda(t) = 
    \begin{cases}
    -\lambda, \quad \quad & 0\leqslant t < \frac{T}{2} \\
    \lambda, \quad & \frac{T}{2} \leqslant t < T
    \end{cases}.
\end{equation}
We have made explicit an energy scale $\Omega$ associated with $H_{PXP}$. Using the spin-1 representation, the Hamiltonian projected to the scar subspace takes the approximate form:
\begin{equation}
    P_{\mathcal{W}} H(t) P_{\mathcal{W}} \approx \sqrt{2} \Omega \sum_i S_i^x + \lambda(t) \sum_i S_i^z,
\end{equation}
and the projected Floquet unitary:
\begin{equation}
    P_{\mathcal{W}} U_F P_{\mathcal{W}} \approx e^{-i (\sqrt{2}\Omega \sum_i S_i^x + \lambda \sum_i S_i^z) (T/2)} \ e^{-i (\sqrt{2} \Omega \sum_i S_i^x - \lambda \sum_i S_i^z) (T/2)}.
\end{equation}
We require that the above equation equals $e^{-i \hat{{\bm n}}\cdot {\bm S} \phi}$ for some rotation axis $\hat{{\bm n}}$ and angle $\phi$. Expand the above equation for the Floquet unitary using the following identity for the spin-1 operators:
\begin{equation}
    e^{-i \hat{{\bm n}}\cdot {\bm S} \theta} = \mathbb{1} + (\hat{{\bm n}} \cdot {\bm S})^2 ({\rm cos}\theta -1 ) - i{\rm sin}\theta (\hat{{\bm n}}\cdot {\bm S}),
\end{equation}
we find
\begin{equation}
    {\rm cos}\phi = 2({\rm cos}\varphi -1) - {\rm cos}2\theta \ {\rm sin}^2\varphi + (2{\rm sin}^4\theta - 2{\rm sin}^2\theta +1)({\rm cos}\varphi-1)^2+1,
\end{equation}
where
\begin{equation}
    \varphi \equiv \frac{1}{2}\sqrt{2(\Omega T)^2 + (\lambda T)^2}, \quad {\rm tan}\theta = \frac{\lambda}{\sqrt{2}\Omega}.
\end{equation}

\begin{figure}[hbt]
\includegraphics[width=0.45\textwidth]{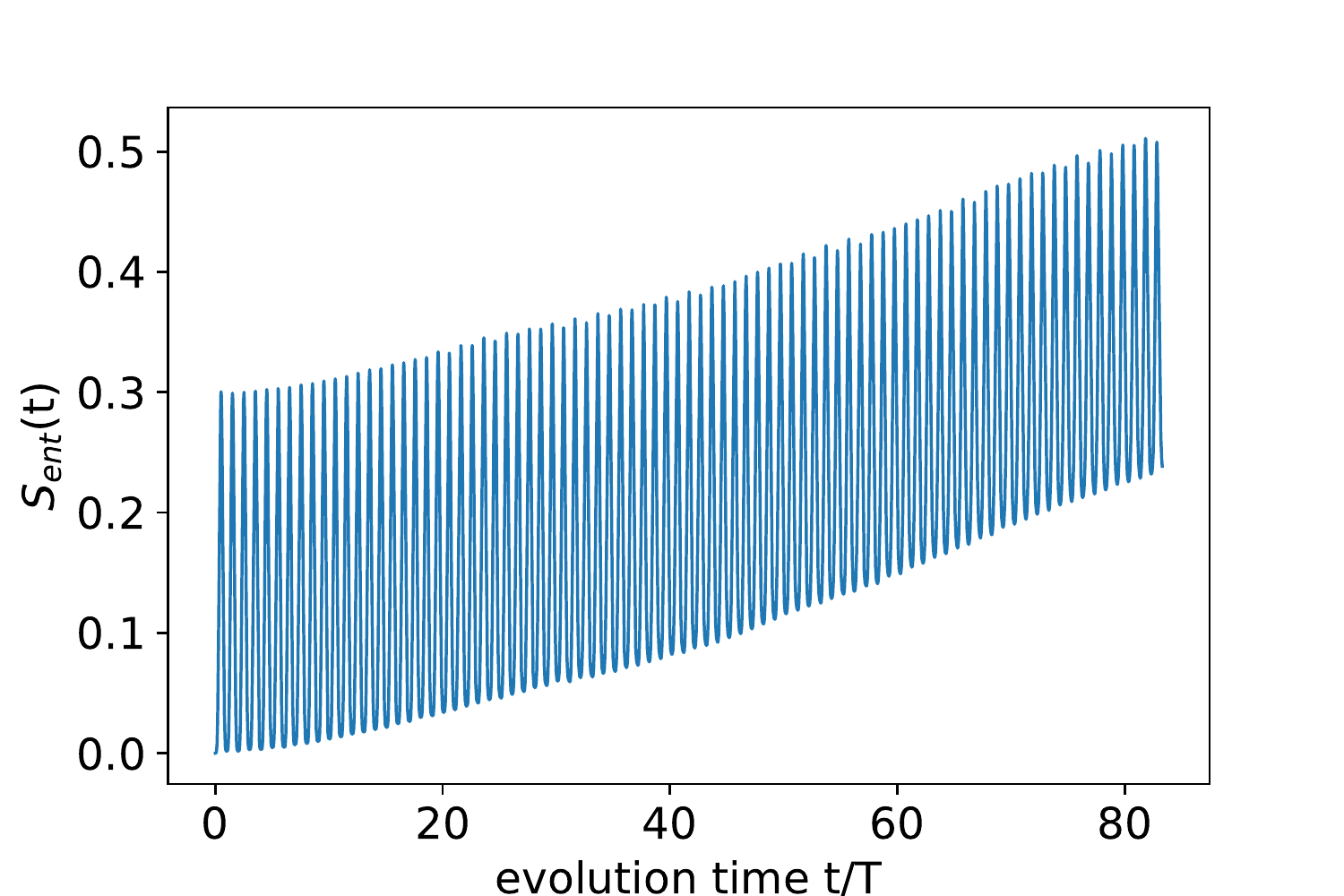}
\caption{Entanglement entropy evolution of the driven deformed PXP model at longer times. Choices of parameters and initial states are the same as Fig.3 in the main text.}
\label{fig:EE_pxp} 
\end{figure}

In Fig.~\ref{fig:EE_pxp}, we plot the evolution of the entanglement entropy under the driven deformed PXP model at longer times, starting from the N\'eel initial state. We find that the entanglement entropy oscillates around a mean value that slowly increases with time. Moreover, the amplitude of the oscillations does not decay at later times. This behavior is a consequence of an enhanced su(2) dynamics with a small leakage outside the subspace $\mathcal{W}$.

\section{The spin-1 $XY$ model}

In this section, we apply the general scheme discussed in the main text to the spin-1 $XY$ model studied in Ref.~\cite{PhysRevLett.123.147201}, and demonstrate that its appropriately driven version can similarly exhibit period-doubled dynamics. The Hamiltonian without the drive is given by:
\begin{eqnarray}
    H_{XY} &=& J\sum_{i}(S_x^iS_{i+1}^x+S_i^yS_{i+1}^y)+h\sum_iS_i^z+D\sum_i\left(S_i^z\right)^2\nonumber \\ &\equiv& H_A + h\sum_iS_i^z+D\sum_i\left(S_i^z\right)^2
\end{eqnarray}
Here we consider the 1D case. It was shown in Ref.~\cite{PhysRevLett.123.147201} that this model hosts an exact tower of scar states created by repeated actions of a ladder operator
\begin{equation}
     J^{\pm} = \frac{1}{2}\sum_i(-1)^i\left(S_i^{\pm}\right)^2
\end{equation}
on a fully polarized reference state: $\ket{\mathcal{S}_n} = \mathcal{N}(n)\left(J^{+}\right)^n\ket{\Omega}$, where $\ket{\Omega} = \bigotimes_{i}\ket{m_i = -1}$ and $\mathcal{N}(n)$ is a normalizaiton factor. One can check that the scar states are annihilated by $H_A$: $H_A |\mathcal{S}_n\rangle = 0.$ Furthermore, the scar subspace is invariant under an su(2) algebra generated by $J^{\pm}$ and
\begin{equation}
    J^z = \frac{1}{2}\sum_i S_i^z,
\end{equation}
although the Hamiltonian itself does not have an SU(2) symmetry. The term proportional to $J^z$ lifts the degeneracy within the scar subspace, and hence this model also falls into the general class discussed in the main text.


Now we can define another generator of the su(2) algebra
\begin{eqnarray}
    J^x &=& \frac{1}{2}(J^{+}+J^{-}) \nonumber \\
    &=& \frac{1}{4}\sum_i(-1)^i\left[(S_i^{+})^2 + (S_i^{-})^2 \right]\nonumber \\
    &=& \frac{1}{2} \sum_i (-1)^i \left[(S_i^x)^2 - (S_i^y)^2 \right].
\end{eqnarray}
Making use of the relation $(S_i^x)^2 + (S_i^y)^2 + (S_i^z)^2 = {\bm S}_i^2 = S(S+1)$ with $S=1$, we can rewrite the above expression as
\begin{equation}
    J^x = \frac{1}{2}\sum_i (-1)^i\left[ 2(S_i^x)^2 + (S_i^z)^2 - S(S+1) \right].
\end{equation}
Notice that in the scar subspace, the spin state of each site is either in state $|m_i=-1\rangle$ or $|m_i=+1\rangle$, and hence $(S_i^z)^2$ is a constant in this subspace. Therefore, it suffices to couple the drive to the operator $\sum_i (-1)^i (S_i^x)^2$.

Consider the time-periodic Hamiltonian:
\begin{equation}
    H(t) = H_{XY} + \lambda(t)\sum_i(-1)^i \left(S_i^x\right)^2
    \label{eq:spin1_drive}
\end{equation}
with
\begin{equation}
    \lambda(t) = 
    \begin{cases}
    \lambda, \quad \quad & 0\leqslant t < \frac{T}{4} \\
    -\lambda, \quad & \frac{T}{4} \leqslant t < \frac{3T}{4}  \\
    \lambda, \quad   & \frac{3T}{4} \leqslant t < T
    \end{cases}.
\end{equation}
The Hamiltonian projected to the scar subspace $\mathcal{W}$ takes the form:
\begin{equation}
    P_{\mathcal{W}} H(t) P_{\mathcal{W}} = 2hJ^z + \lambda(t)J^x.
\end{equation}
 Here we dropped the constant terms which do not affect the Floquet evolution in the scar subspace. Notice that since $(S_i^z)^2=1$ is a constant in the scar subspace, a non-zero value of $D$ does not affect the physics. The corresponding Floquet unitary can be obtained similarly:
\begin{equation}
    P_{\mathcal{W}} U_F P_{\mathcal{W}} = e^{-i (2hJ^z + \lambda J^x) (T/4)} \ e^{-i (2h J^z - \lambda J^x) (T/2)} e^{-i (2hJ^z + \lambda J^x) (T/4)}.
\end{equation}
From the discussion above, we see that the generators $J^z$ and $J^x$ can be decomposed to on-site operators: $J^z = \sum_i J^z_i$, $J^x = \sum_i J^x_i$, and these on-site operators also constitute the corresponding su(2) algebra on a single site. Hence, the effective rotation angle and rotation axis can be obtained following the same calculations as in Appendix~\ref{sec:PXP}.

\begin{figure}[hbt]
  \centering
  \subfigure[]{
    \includegraphics[width=0.33\textwidth]{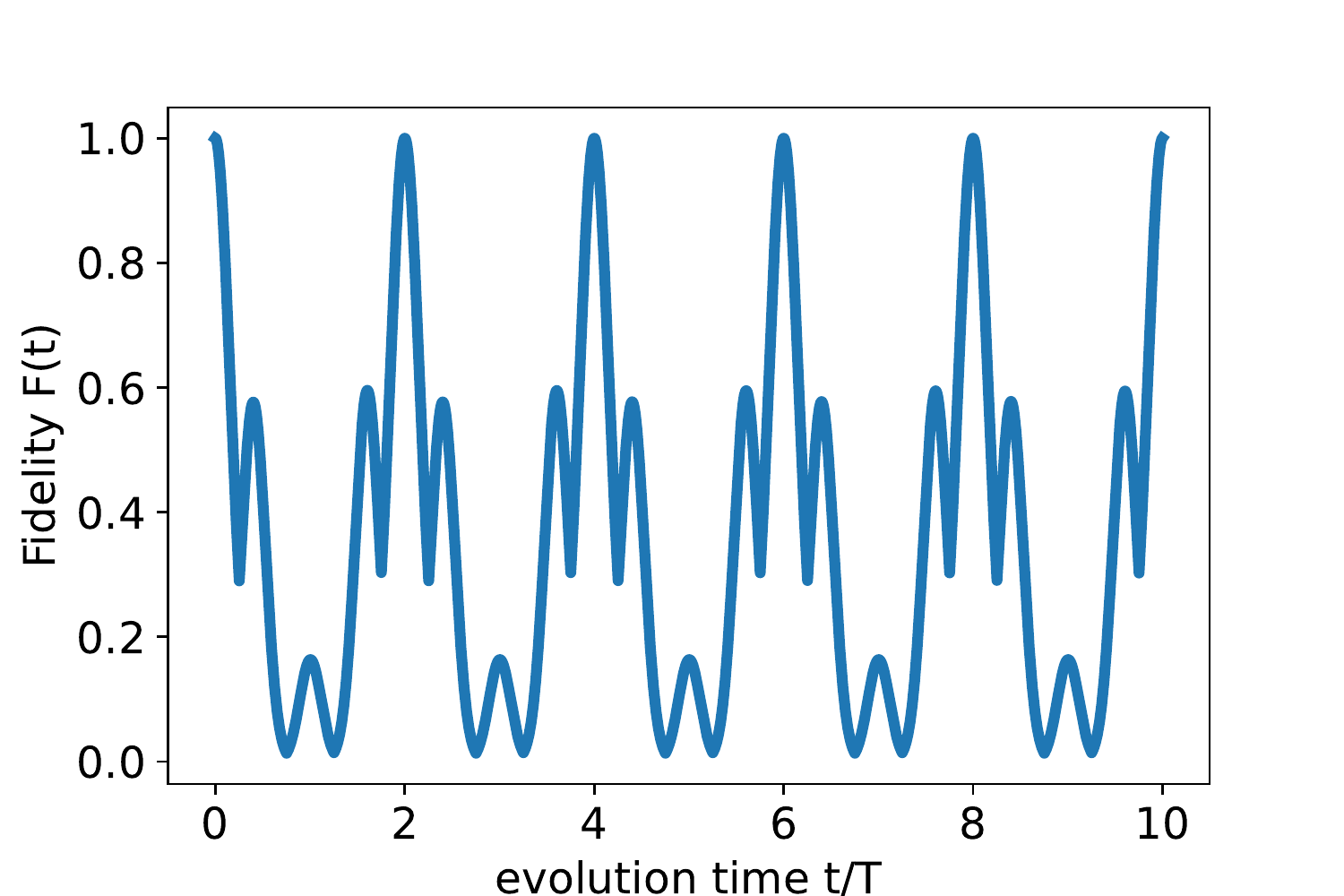}
  }%
  \subfigure[]{
    \includegraphics[width=0.33\textwidth]{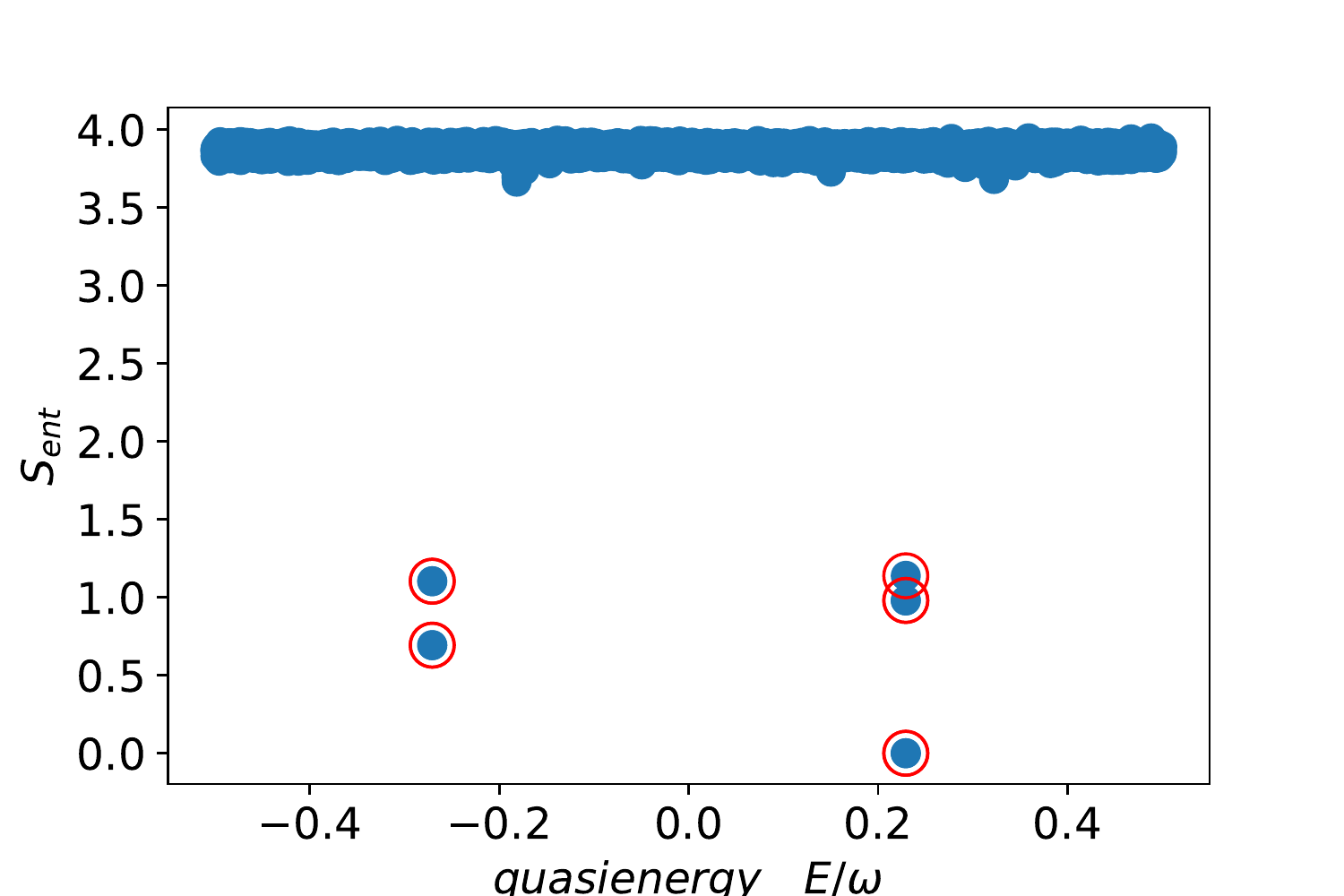}
  }%
  \subfigure[]{
    \includegraphics[width=0.33\textwidth]{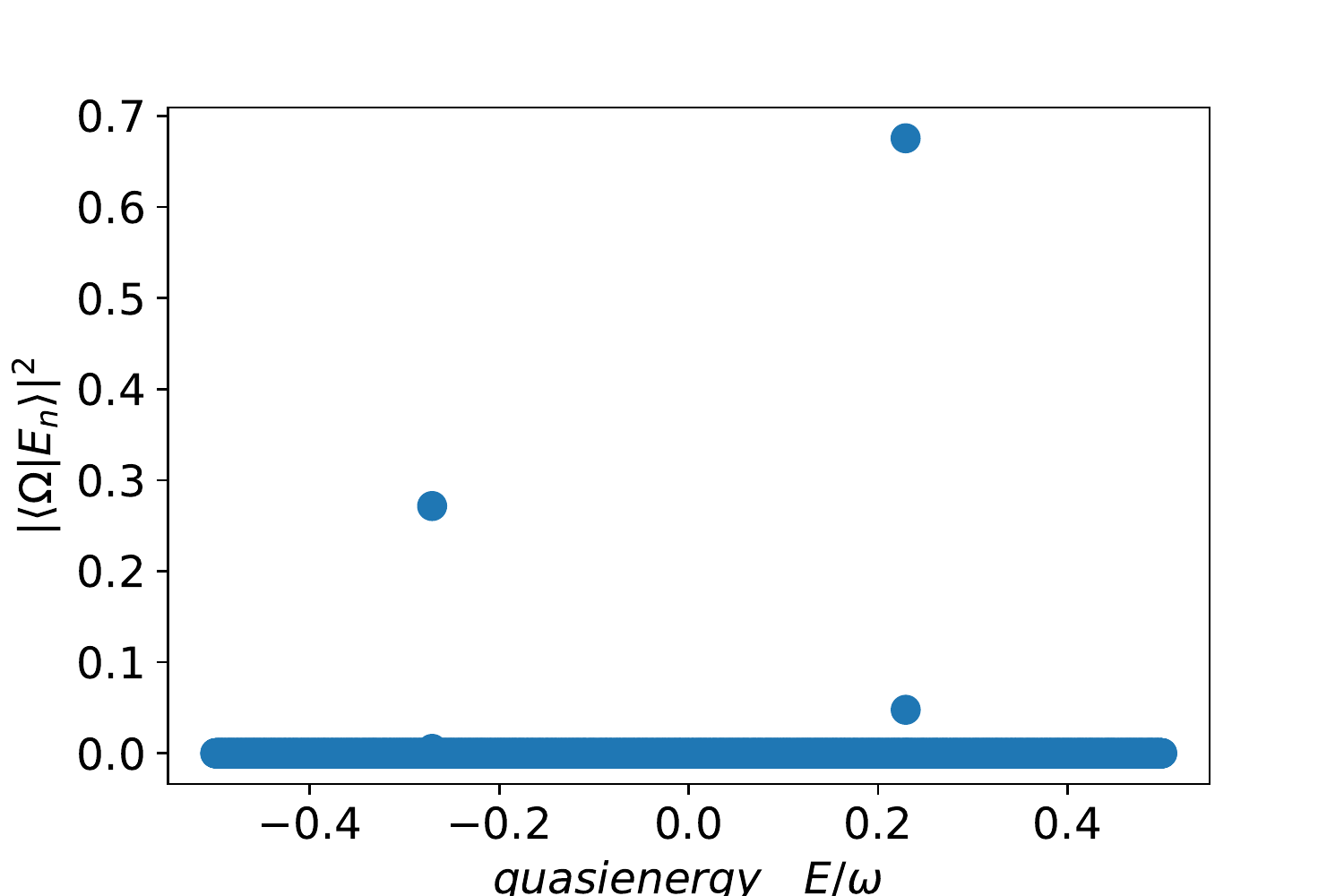}
  }
  \caption{Numerical results for the driven spin-1 model~(\ref{eq:spin1_drive}). (a) The fidelity $F(t)$ exhibits perfect revivals at integer multiples of $2T$. (b) The quasienergy spectrum features a set of weakly-entangled eigenstates with quasienergy spacing $\pi$ between them. (c) Overlap of the initial state $|\Omega\rangle= |- - \cdots -\rangle$ with the Floquet eigenstates. Results are obtained for system size $L=8$, $h T \approx 1.8 $, $\lambda T \approx 3.0132$, and $D= 0.1$. }
  \label{fig:spin1}
\end{figure}

Below, we show numerical results for the driven spin-1 $XY$ model. We find that for appropriately chosen driving parameters such that $\phi=\pi$, the fidelity indeed exhibits perfect revivals at twice the period of the drive, as shown in Fig.~\ref{fig:spin1}(a). This period-doubled dynamics is again due to the emergence of $\pi$-paired Floquet eigenstates with low entanglement, with which the initial state $|\Omega \rangle = |--\cdots -\rangle$ overlaps [Fig.~\ref{fig:spin1}(b)\&(c)].

\end{document}